\documentclass{elsart}
\usepackage{natbib}
\usepackage{epsfig}
\begin{document}
\runauthor{Cicero, Caesar and Vergil}
\title{
Fragmentation Production of $\Omega_{ccc}$ and $\Omega_{bbb}$
Baryons}
\author{M.A. Gomshi Nobary}
\address{Department of Physics, Faculty of Science, Razi University,Kermanshah, Iran.}
\address{The Center for Theoretical Physics and Mathematics, A.E.O.I.,
Roosbeh Building, P.O. Box 11365-8486 Tehran, Iran. }

\noindent E-mail:mnobary@aeoi.org.ir

\vskip .6 cm \noindent {\bf Abstract}

The $\Lambda$ baryons with a single heavy flavor which transfer
the quark polarization, have been studied both theoretically and
experimentally. The $\Xi$'s with two heavy constituents are well
treated in quark-diquark model. In this work we study the
production of triply heavy baryons in the perturbative QCD regime
and calculate the fragmentation functions for $\Omega_{ccc}$ and
$\Omega_{bbb}$ in the $c$ and $b$ quark fragmentation,
respectively. We then obtain the total fragmentation probability
and the average fragmentation parameter for each case.


\section {Introduction}
The quark model of hadrons has proved to be successful in
describing hadrons and their properties. In the heavy quark
sector it predicts hadrons having $c$, $b$ and $t$ quarks as
constituents. However, the discovery of the top quark [1] and
the determination of its lifetime [2] made it clear that it
cannot participate in strong interactions and therefore only the $c$
and $b$ flavors are left to take part in the hadron production
interplay.

Recently meson states constituting heavy flavor have received
considerable attention. Specially $B_c$ and $B_c^*$ states with
$\overline b c$ quark content have been in focus both
theoretically [3] and experimentally [4] in the last few years. It
is established that the fragmentation functions describing their
production mechanism are calculable in perturbative QCD [3] and
hence the total fragmentation probabilities and production cross
sections are calculated in due course.

Baryon states with heavy flavor fall into three categories. States
containing one heavy flavor such as $\Lambda_c$ and $\Lambda_b$
are interesting states due to the fact that they carry the
original heavy flavor polarization. They are presently being studied
experimentally [5]. The second category involves
baryons with two heavy flavor like the states $\Xi_{cc}$,
$\Xi_{bb}$ and $\Xi_{bc}$ [6]. They are treated within the
approximate quark-diquark model [7]. The model treats the
production of the so called diquark perturbatively similar to the
states such as $B_c$. Then, it can be proved that the formation of a baryon
out of the diquark is almost the same as the fragmentation of an
antiquark into a meson. In this way one obtains the fragmentation
functions, the total production probabilities and other
relevant parameters which specify their properties. In the third
category, we have baryons with three heavy constituents. If we
follow the scheme used in the case of heavy mesons and assume that
their fragmentation functions are calculable in the perturbative
regime, then we can calculate Feynman diagrams like the one in
figure 1 to obtain the fragmentation functions. There are eight
such diagrams in the lowest order contributing triply heavy baryons,
i.e. $\Omega_{ccc}$, $\Omega_{ccb}$, $\Omega_{cbb}$ and
$\Omega_{bbb}$ production [8].

In this paper our aim is to calculate the fragmentation of the $\Omega_{ccc}$
and $\Omega_{bbb}$ baryons in the lowest order perturbative regime and obtain
their fragmentation functions in an exact analytical form.

\section {Kinematics }
We consider the fragmentation of a heavy quark $Q$ into a $QQQ$
system with three identical flavor. This procedure is illustrated
in Figure 1. We have used an infinite momentum frame in which all
of the particles are moved in the forward direction, i.e. the
longitudinal direction along the $z$ axes, where the $QQQ$ moves.
We let the original quark keep its transverse momentum.
Furthermore, we assume that the two antiquark jets move almost in
the same direction. This assumption is justified due to the fact
that the very high momentum of the initial heavy quark will
predominantly be carried in the forward direction. Due to momentum
conservation, the total transverse momentum of the two jets will
be identical to the transverse momentum of the initial quark. In
this context the four momenta of the particles will assume the
following form

\begin{figure}
\includegraphics[width=14cm]{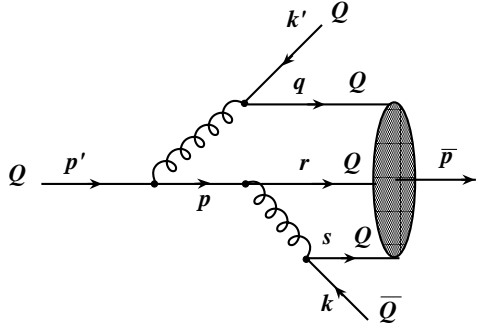}
\caption[Submanifold]{Feynman diagram illustrating the lowest
order fragmentation of a heavy quark, $Q$, into a $\Omega_{QQQ}$
baryon. The four momenta are labelled.}
\label{f.emb}
\end{figure}

\begin{eqnarray}
p'_\mu &=[p'_\circ,{\bf p'}_T,p'_L],\;q_\mu=[q_\circ,{\bf
0},q_L],\;r_\mu =&[r_\circ,{\bf 0},r_L]\nonumber\\
s_\mu &=[s_\circ,{\bf 0},s_L], \;k_\mu=[k_\circ,{\bf
k}_{T},k_L],\; k'_\mu =&[k'_\circ,{\bf k'}_{T},k'_L].
\end{eqnarray}

We have used the fragmentation parameter, $z$, as defined in the literature, i.e.,

\begin{eqnarray}
z={{(E+p_\Vert)_B}\over {(E+p_\Vert)_Q}}
={{E_B}\over {E_Q}}.
\end{eqnarray}

\noindent The last step follows form application of the infinite momentum
frame. Therefore, the final state particle energies are parameterized as follows

\begin{eqnarray}
\overline p_\circ=zp'_\circ,
\end{eqnarray}

\noindent where $\overline p_\circ =r_\circ+q_\circ+s_\circ$ is the energy
of the baryon. Therefore

\begin{eqnarray}
r_\circ=x_1zp'_\circ,\;\;\;\;
q_\circ=x_2zp'_\circ,\;\;\;\;
s_\circ=x_3zp'_\circ.
\end{eqnarray}

\noindent Here the $x$'s are the energy ratios carried by the constituents.
Since the constituents are identical and fly together, it is found
that $x_1=x_2=x_3=1/3$. This is consistent with our argument about the wave
function for such states in the next section. We also have assumed that the
two anti-quarks which
initiate the two jets have equal energies, i.e.
\begin{eqnarray}
k_\circ=k'_\circ={1\over 2}(1-z)p'_\circ.
\end{eqnarray}

\noindent On the other hand due to our discussion about transverse momentum, we have

\begin{eqnarray}
{\bf k}_T={\bf k'}_T={1\over 2}{\bf p'}_T.
\end{eqnarray}

\noindent We will discuss this later assumption in the final section.

\section{ Calculation of the Fragmentation Functions for $\Omega_{QQQ}$ }

We are now ready to calculate the diagram shown in figure 1.
The fragmentation of a heavy quark $Q$ into a heavy
baryon $\Omega_{QQQ}$ is obtained by squaring the total amplitude and
integrating over final state phase space,

\begin{eqnarray}
D_Q^B(z,\mu_\circ)={1\over 2}\sum_s \int \vert T_B
\vert^2\delta^3( \overline{{\bf p}}+{\bf k}+{\bf k'}-{\bf p'})
{\rm d^3 \overline{{ p}}}{\rm d^3 { k}}{\rm d^3 { k'}},
\end{eqnarray}

\noindent where $T_B$ is the amplitude of the baryon production
which involves the hard scattering amplitude $T_H$ and the
non-perturbative smearing of the bound state. The average over initial
spin states and the sum over final spin states are performed. The
heavy hadron production amplitude is composed of a partonic part,
which can be calculated using perturbative QCD, and a
non-perturbative part, which describes the transition of free
quarks into the final state hadron. In the framework of
non-relativistic quark model, this non-perturbative part could be
accounted for through the wave function which is calculable using
potential models. Since at present there is no known information
concerning such wave functions, we have assumed a delta
function type wave function for them. This assumption guarantees
that the constituents will fly parallel and have no transverse
momentum with respect to their direction of motion. This is also
consistent with our assumptions in section 2. The hard
scattering amplitude which is obtained by perturbative
calculations of the tree diagram in figure 1, may be put in the
following form [9],

\begin{eqnarray}
T_H={{24\pi^2\alpha_s^2 m^4C_F} \over {\sqrt{2p'_\circ \overline
p_\circ k_\circ k'_\circ}}} {\Gamma \over
{g_1(z)g_2(z)g_3(z)(\overline
p_\circ+k_\circ+k'_\circ-p'_\circ)}}.
\end{eqnarray}

\noindent Here $\alpha_s=g^2/4\pi$ is the strong interaction
coupling constant and $\Gamma$ indicates that part of the
amplitude which embeds spinors and gamma matrices. The $1/g$'s are
the propagators of the two gluons and the intermediate fermion
respectively.

To absorb the soft behavior of the bound state into hard scattering
amplitude we have used the scheme introduced in [7].
The probability amplitude at large momentum transfer
 factories into a convolution of the hard-scattering amplitude $T_H$,
and baryon-distribution amplitude $\phi_M$ [10], i.e.,

\begin{eqnarray}
T_B(k_i,p_i)=\int\bigl[ dx \bigr] T_H(k_i,p_i,x_i) \phi_B
(x_i,q'^2),
\end{eqnarray}

\noindent where $T_H$ is given by (8) and $\phi_B$ is the
probability amplitude to find quarks co-linear up to a scale
$q'^2$ in the baryonic bound state. In (9), $x_i$'s are the
momentum fractions carried by the constituent quarks and $\bigl[
dx \bigr]=dx_1dx_2dx_3 \delta(1-x_1-x_2-x_3).$ In view of our
early discussion in this section, we propose the following
expression for the probability amplitude

\begin{eqnarray}
\phi_B= {f_B}\; \delta\biggl\{x_i-{m_i\over m_B}\biggr\},
\end{eqnarray}

\noindent where $m_B$ is the baryon mass and $f_B$ refers to the
characteristics of the baryon bound state and is similar to the
meson bound state where the decay constant $f_M$ is introduced.
Putting this expression and (8) in (9) and carrying out the
necessary integrations, we find

\begin{eqnarray}
T_B={{24\pi^2\alpha_s^2m^4f_BC_F}\over {\sqrt{2p'_\circ \overline
p_\circ k'_\circ k_\circ}}} {\Gamma \over
{g_1(z)g_2(z)g_3(z)(\overline
p_\circ+k_\circ+k'_\circ-p'_\circ)}}.
\end{eqnarray}

\noindent Now we are able to obtain the fragmentation function in
(7) as
\begin{eqnarray}
D(z,\mu)&=&{{(48\pi^2\alpha_s^2m^{4}f_BC_F)^2}\over {8}}\nonumber\\
&  &\times\int {{\frac{1}{2}\sum_s\overline{\Gamma}\Gamma
\delta^3(\overline {\bf p}+{\bf k}+{\bf k'}-{\bf p'}) {\rm d^3
\overline { p}}{\rm d^3 { k}}{\rm d^3 { k'}}     } \over
{{\overline p_\circ p'_\circ k_\circ
k'_\circ}\Bigl[g_1(z)g_2(z)g_3(z)(\overline
p_\circ+k_\circ+k'_\circ-p'_\circ)\Bigr]^2}}.
\end{eqnarray}

Spin sum-average of $\overline{\Gamma}\Gamma$ for Figure 1 is
easily calculated using the REDUCE. To do the phase space
integrations in (12), first we consider the integral,
\begin{eqnarray}
I&=&\int {{ \delta^3(\overline{\bf p}+{\bf k}+{\bf k'}-{\bf
p'}){\rm d}^3\overline{ p}} \over {p'_\circ(\overline
p_\circ+k_\circ+k'_\circ-p'_\circ)^2}} \nonumber\\
& &=\frac{p'_\circ}{f(z)^2},
\end{eqnarray}

\noindent where

\begin{eqnarray}
f(z)= -{{p'_T}^2\over{3m^2}}+ {3\over z}+ {{4}\over{3}}
\biggl(1+{{p'_T}^2\over {4m^2}}\biggr){1\over{1-z}}.
\end{eqnarray}

\noindent Also we note that
\begin{eqnarray}
\int f(z,{\bf k}_T)d^3 {\bf k}&=&\int f(z,{\bf k}_T){\rm d}k_L
 {\rm d}^2 k_T\nonumber\\
  &  &=m^2 k_\circ f(z,\langle
{k_T}^2\rangle ) =m^2 k_\circ f(z,{1\over 2}\langle
{p'_T}^2\rangle ),
\end{eqnarray}

\noindent and
\begin{eqnarray}
\int f(z,{\bf k'}_T)d^3 {\bf k'}&=&\int f(z,{\bf k'}_T){\rm d}k'_L
{\rm d}^2 k'_T\nonumber\\
& &=m^2 k'_\circ f(z,\langle {k'_T}^2\rangle ) =m^2k'_\circ
f(z,{1\over 2}\langle {p'_T}^2\rangle ).
\end{eqnarray}
\noindent Here, instead of performing transverse momentum
integrations, for simplicity we have replaced them by their
average values. Putting all this back in (12), we obtain the
fragmentation function as,
\begin{eqnarray}
D_{Q\rightarrow
QQQ}(z,\mu_\circ)&=&{{\pi^4\alpha_s^4f_B^2C_F^2}\over {108
m^2z^4(1-z)^4 f(z)^2 g(z)^6}}\nonumber\\
&&\times\biggl[\xi^8 z^8+4\xi^6 z^6(83-130 z+51 z^2)\nonumber\\
&&+6\xi^4z^4(1413-3084z+3022z^2-2156z^3+821z^4)\nonumber\\
&&+4\xi^2 z^2(18711-51678z+69417z^2-70308z^3\nonumber\\
&&+53529z^4-25950z^5+6343z^6)+222345-740664z\nonumber\\
&&+1179036z^2-1253448z^3+90126z^4-388872z^5\nonumber\\
&&+109916z^6-49912z^7+20649z^8\biggr].
\end{eqnarray}

\noindent Here we have defined $\xi=\langle{ p'}_T^2\rangle/m^2$.
$g(z)$ comes from the propagators and have the following form

\begin{eqnarray}
g(z)=1+{3\over z}+ \frac{4}{3}\biggl(1+{{\bf p'}_T^2\over{4
m^2}}\biggr){{z}\over{1-z}} .
\end{eqnarray}
and $f(z)$ is due to the energy denominator given by (14).
Replacement of $f(z)$ by $g(z)$ which we have done in the original
manuscript, changes the fragmentation function only slightly.

The fragmentation function $D_{c\rightarrow
\Omega_{ccc}}(z,\mu_\circ)$ and $D_{b\rightarrow
\Omega_{bbb}}(z,\mu_\circ)$ are easily obtained from the above by
letting $m=m_c,m_b$ and using appropriate $f_B$, $\alpha_s$ and
$\mu_\circ$ values.

\section{Results and Discussion}
We were able to calculate the process of direct $c$ and $b$ quark
fragmentation into $\Omega_{ccc}$ and $\Omega_{bbb}$ baryons. In
doing so we had to follow certain assumptions. Firstly we have
considered only the dominant contributing Feynman diagram in
leading order. This assumption reduced the complexity and the
length of the calculation and enabled us to obtain analytic forms
of the fragmentation functions. Our second assumption concerns
kinematics. We believe that the high momentum of the process has
to be taken away in the forward direction and let the two
antiquarks carry the transverse momentum of the initial heavy
quark. Furthermore since they are identical, we have considered
equal contribution from them both in magnitude an in direction.
Therefore, we have established equations (5) and (6) and used them
in our calculation. To see how our later assumption works, we have
set the kinematics by allowing $k$ and $k'$ to share the jet
energy-momentum. We have let $k=x(1-z)p'$ and $k'=(1-x)(1-z)p'$
where $x$ is a variable which is between zero and one. We have
repeated our calculations and studied the behaviour of
$\Omega_{ccc}$ fragmentation function with the same parameters as
before. It is revealed that as $x$ increases, the function grows
rapidly and gives the highest peak at $x=1/2$. As $x$ increases
further, the peak falls rapidly. Since there is not much
information about the wave functions of the triply heavy baryons
at hand, we have reduced the non-perturbative smearing of the
bound state to a delta function times a factor which is much like
the meson decay constant. We have denoted this constant by $f_B$
and assumed to take 0.25 GeV both for $\Omega_{ccc}$ and
$\Omega_{bbb}$ baryons.

In obtaining (17) we have not performed the transverse momentum
integrations. Instead we have replaced the variables by their
average values. However the numerical integration converges well
for sufficiently large transverse momentum. Let us now sketch the
behaviour of our fragmentation functions. Figure 2 shows the
behaviour of $D_{c\rightarrow \Omega_{ccc}}(z,\mu_\circ)$ and
$D_{b\rightarrow \Omega_{bbb}}(z,\mu_\circ)$ in the fragmentation
scale $\mu_\circ$. In drawing them we have assumed that $m_c=1.25$
GeV and $m_b=4.25$ GeV. The scales are $\mu_\circ$=6.25 GeV and
$\mu_\circ$=21.25 GeV respectively. We have set $\langle
{p'_T}^2\rangle$= 1 GeV and included the colour factor of
$C_F$=7/6 obtained using color line counting rule. Consistent with
the study of $B_c$ and $B_c^*$ states, we have taken
$\alpha_s=0.26 $ for $\Omega_{ccc}$ and $\alpha_s=0.18$ for
$\Omega_{bbb}$ [3].

\begin{figure}
\hskip 4cm\includegraphics[width=8cm]{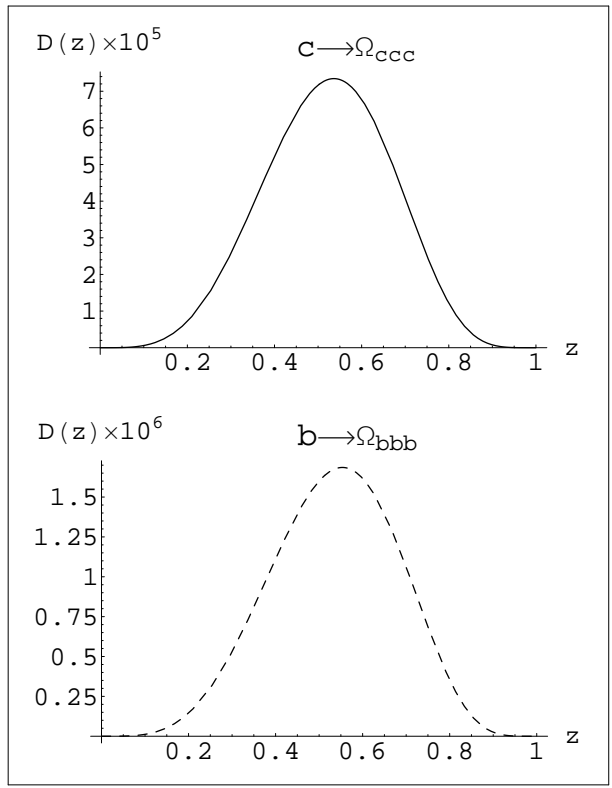}
\caption[Submanifold]{The behavior of  $\Omega_{ccc}$ (solid) and
$\Omega_{bbb}$ (dashed) fragmentation function at the respective
fragmentation scale.} \label{f.emb}
\end{figure}

At leading order in $\alpha_s$ one has $\int_0^1 P_{Q\rightarrow
Q}(z,\mu){\rm d}z=0$ [12], and the evolution equation implies that
the fragmentation probability $\int_0^1 D_{Q\rightarrow
B}(z,\mu){\rm d}z$ does not evolve with the scale $\mu$.
Therefore, the fragmentation probability is a universal
characteristic of the production rates. The evolution only moves
the $z$-distribution to small values of $z$. We have obtained this
quantity for $\Omega_{ccc}$ and $\Omega_{bbb}$ using our
fragmentation functions. The other relevant kinematical parameter
is the average fragmentation parameter. Our results for the
fragmentation probabilities and $\langle z\rangle$ appear in Table
1. It is seen that our analysis give very close $\langle z\rangle$
values for $\Omega_{ccc}$ and $\Omega_{bbb}$. The fragmentation
probabilities in Table 1 suggest that considerable event rate is
expected both at the Tevatron and the LHC .
\begin{table}
\caption{Fragmentation probability and $\langle z \rangle$ for different states.}
\begin{tabular}{l l l}
\hline
  & $\Omega_{ccc}$& $\Omega_{bbb}$\\
\hline
 Frag. Prob.& $2.789\times10^5 $&$6.459\times10^7 $ \\
$\langle z \rangle$ &0.522 & 0.535\\
\hline
\end{tabular}
\end{table}

\end{document}